# Assessing the Precision and Recall of msTALI as Applied to an Active-Site Study on Fold Families

Devaun McFarland
Department of Computer science and Engineering
(University of South Carolina)
Columbia, SC 29208 USA
mcfarlad@email.sc.edu

Homayoun Valafar
Department of Computer science and Engineering
(University of South Carolina)
Columbia, SC 29208 USA
homayoun@cec.sc.edu

*Abstract*—Proteins execute various activities required by biological cells. Further, they structurally support and promote important biochemical reactions which functionally are sparked by active-sites. Active-sites are regions where reactions and binding events take place directly; they foster protein purpose. Describing functional relationships depends on factors that incorporate sequence, structure, and the biochemical properties of amino acids that form proteins. Our approach to active-site description is computational, and many other approaches utilizing available protein data fall short of ideal. Successful recognition of functional interactions is crucial to advancements in protein annotation and the bioinformatics field at large. This research outlines our Multiple Structure Torsion Angle Alignment (msTALI) as a suitable strategy for addressing active-site identification by comparing results to other existing methods. Specifically, we address the precision of msTALI across three protein families. Our target proteins are PDBIDs 1A2B, 1B4V, 1B8S, 1COY, 1CXZ, 3COX, 1D7E, 1DPF, 1F9I, 1FTN, 1IJH, 1KOU, 1NWZ, 2PHY, and 1SIC.

*Keywords—msTALI, active-sites, function, precision, recall*

I. INTRODUCTION

Identification of protein active-sites is pivotal for better understanding their function. An active-site is a region of a protein where binding to its substrate is facilitated and it therefore describes a protein's function. Given the plethora of available structural information on proteins, new methods for the discovery of active-sites should be plausible. For instance, if proteins demonstrate similar function, then there must be mirroring structural similarities. Capturing the direct relationship between sequence, structure, and function of a protein is a complex problem that has not been fully understood. This task is difficult since such relationships require accurate descriptions that are not always classified or recognizable. Several factors contribute to functionality of a protein such as: the location and size of active-sites, ligand binding properties, and regions of proteins that are surface accessible [1]. Further, it is not known which factors are most responsible for description. With this, the inherent problem is development of an automated methodology that successfully identifies binding regions, while incorporating any additional requisites for the specific enzymatic activity. Impacts of such methods build our understanding of the molecular basis for diseases, drug design, targeting mutants, functional annotation for unknown proteins, and for studies in protein design and engineering.

To start protein surfaces are irregular and docking techniques are utilized to explore interactions [2]. Typically, geometric approaches are used to locate active-sites by mapping a protein's surface space, from which, grids are used [3]. Next cavity/ cleft regions are ranked, categorized, and examined [4, 5]. It is evident that cavity features are relevant [6] as are the graph theoretic methods that incorporate hashing techniques [7]. Neural Networks (NN) have also been employed for comparing the structure function similarities phenomenon [8]. Still, training for protein interactions is complex so fuzzy functional forms (FFFs) are adopted to strengthen various approaches to locate active-sites [9][10]. Collectively, all these computational methods have purpose and have throttled the common core. There are even web services aiming to address active-sites [11][12][13].

The comparison methods discussed in this paper represent some of the most recognizable methodologies for active-site identification to date. In fact, each employs several of the grid based, surface mapping, and detection techniques mentioned above. SiteEngine is a recognized method for pairwise docking descriptions with hash triangles [14]. SuMo incorporates chemical groups with structural representations [12] and pdbFun is a web service that breaks down its analysis at the residue level [15]. Binding Site Finder (BsFinder) methodology provides a three step process similar to our goal since it incorporates sequence and structure information [16].

Though, limitations for many preexisting approaches of active-site identification are prevalent. Our Multiple Structure Torsion Angle Alignment (msTALI) approach addresses the shortcomings in other approaches by incorporating a multitude of properties for groups of proteins simultaneously [17]. This integrated approach is also dynamic, just as proteins are [18], and generates competitive results. Further, by directly comparing msTALI to BsFinder – based on it outperforming other methods, mentioned in the next section – we discuss precision, recall, and how it measures to standards.

## II. BACKGROUND AND METHODS

In this study, we evaluate the performance of our approach in identifying the active-site of 15 proteins previously studied by other methods. The following sections provide an overview of the previous related work, a more detailed description of the target proteins and our approach.

### A. Target proteins

Our structure-based identification of active-sites relied on analysis of 15 proteins listed in Table 1. The selection of these proteins was based on the existence of previous reported results from other methods to which we can compare our results [16]. The proteins are classified to three family of enzymatic activities: G proteins family in P-loop folds, PYP-like family in Profilin-like folds, and FAD-linked reductases family in FAD/NAD(P)-binding folds, that are respectively listed in rows 1-5; colored purple, 6-10; colored green, and 11-15; colored red, of the Table 1. The G-domain and Ras superfamily are well known [19], profilin is widely studied for cellular activity [20], and the same holds for analysis of FAD based proteins [21]. The first column in Table 1 alphabetically list the PDB IDs [22] with respect their color coded fold family. The second column in Table 1 lists the organism from which the target protein was selected. It is noteworthy the diversity of organisms within each enzymatic group, a testimony to the broad evolutionary selection of these enzymatic activities. The third column in Table 1 lists the primary binding molecules used to aid in classifying each group of proteins. The last column in Table 1 provides information regarding other binding sites that accommodate binding to co-factor molecules. It is important to note that it can be argued that some proteins may have more than one "active-site" that enables the enzymatic activity of the proteins; one site that binds and facilitates the alteration of the ligand, and the other sites that help enable/disable or regulate the modification of the enzyme.

### B. Structural similarity of the target proteins

With Fig. 1 we illustrate the strong structural similarity for our target proteins across each fold family. The msTALI structural alignment superimposes the proteins in P-loop folds, Profilin-like folds, and FAD/NAD(P)-binding folds, with 173, 119, and 495 conserved residues with backbone RMSD values of 1.30, 0.43, and 0.69 angstroms respectively. Considering the average length of each corresponding protein fold family is roughly, 184, 124, and 505 amino acids long; our proteins are indeed structurally similar.

Table 2 uses the same conventions as Table 1 to highlight some relevant structural properties for the target proteins. The second and third columns of Table 2 provide the size of each protein reported in the number of amino acids, and the structure/sequence classification reported by CATH [23]. CATH classification [23] provides a hierarchical classification of proteins based on Class, Architecture, Topology and Homology (sequence). In short, CATH describes sequence and structural makeup of proteins, which are important in better understanding the protein sequence-structure-function relationship.

As it can be noted from Table 2, proteins from the same enzymatic groups share similar sizes and CATH classification. Fig. 1, exemplifies the structural similarity. While this observation is generally true to some degree, the relationship between structure and function is more diverse than portrayed in this table. Therefore, it is important to note that the similar structures increase the difficulty of our approach as it pertains to active-site description.

*Table 1.* Target proteins described by organism, the ligands and metal complexes they bind. *1. Brevibacterium Sterolicum *2. Halorhodospira halophila

| PROTEIN | ORGANISM | LIGANDS | METALS |
|---|---|---|---|
| 1A2B | Homo Sapiens | GSP | MG |
| 1CXZ | Homo Sapiens | GSP | MG |
| 1DPF | Homo Sapiens | GDP | N/A |
| 1FTN | Homo Sapiens | GDP | MG |
| 1S1C | Homo Sapiens | GNP | MG |
| 1D7E | H. Halophila[*2] | HC4 | N/A |
| 1F9I | H. Halophila[*2] | HC4 | N/A |
| 1KOU | H. Halophila[*2] | DHC (NBU) | N/A |
| 1NWZ | H. Halophila[*2] | HC4 | N/A |
| 2PHY | H. Halophila[*2] | HC4 | N/A |
| 1B4V | Streptomyces sp | FAD | N/A |
| 1B8S | Streptomyces sp | FAD | N/A |
| 1COY | B. Sterolicum[*1] | FAD (AND) | N/A |
| 1IJH | Streptomyces sp | FAD | N/A |
| 3COX | B. Sterolicum[*1] | FAD | N/A |

*Table 2.* Target proteins are listed with their size information in residue length, and with their CATH classifications. C-class A-architecture T-topology H-homology

| PROTEIN | Length (Res) | CATH CLASS |
|---|---|---|
| 1A2B | 182 | 3.30.505.10 |
| 1CXZ | 182 | 3.40.50.300, 1.10.287.160 |
| 1DPF | 180 | 3.40.50.300 |
| 1FTN | 193 | 3.40.50.300 |
| 1S1C | 183 | 3.40.50.300 |
| 1D7E | 122 | 3.30.450.20 |
| 1F9I | 125 | 3.30.450.20 |
| 1KOU | 125 | 3.30.450.20 |
| 1NWZ | 125 | 3.30.450.20 |
| 2PHY | 125 | 3.30.450.20 |
| 1B4V | 504 | 3.50.50.60, 3.30.410.10 |
| 1B8S | 504 | 3.50.50.60, 3.30.410.10 |
| 1COY | 507 | 3.50.50.60, 3.30.410.10 |
| 1IJH | 504 | 3.50.50.80, 3.30.410.10 |
| 3COX | 507 | 3.50.50.60, 3.30.410.10 |

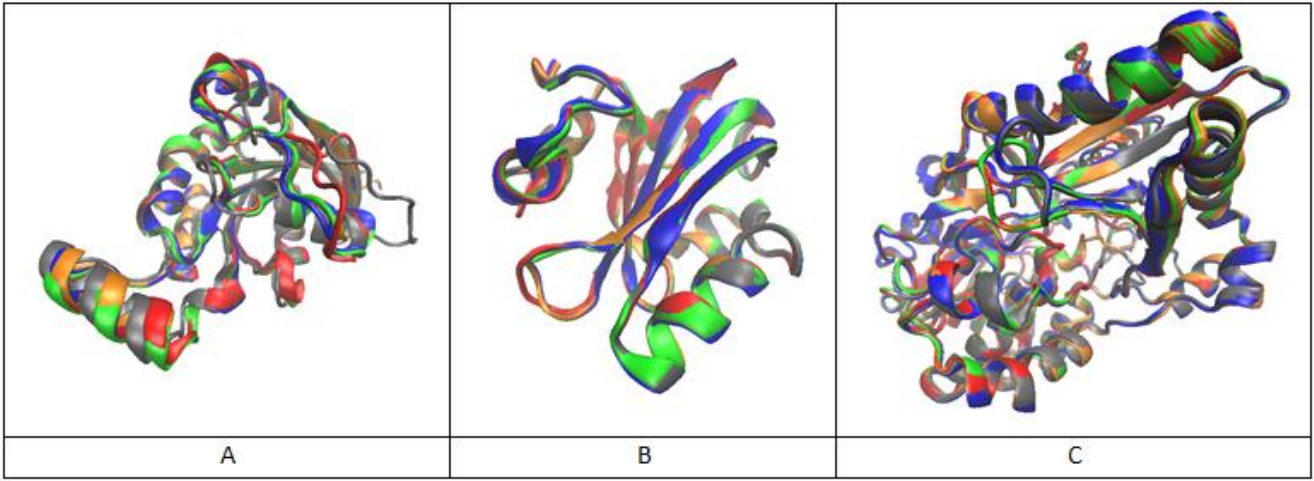

*Figure 1.* Super Imposition of Protein Fold Families. A. Superimposed structures for Proteins 1A2B (green), 1CXZ (orange), 1DPF (grey), 1FTN (red), and 1SIC (blue) from the G proteins family in P-loop folds. B. Proteins 1D7E (green), 1F9I (orange), 1KOU (grey), 1NWZ (red), and 2PHY (blue) form the PYP-like family in Profilin-like folds. C. Proteins 1B4V (green), 1B8S (orange), 1COY (grey), 1IJH (red), and 3COX (blue) from the FAD-linked reductases family in FAD/NAD(P)-binding folds.

*C. Previous Results*

Before proceeding to summarize the previous work that shaped our study, precision and recall have to be defined. For the comparison study *precision* values characterize the accuracy of the program. Precision values are obtained by using the number of sites identified by the program that are confirmed in SitesBase [24] divided by the total number of sites identified by the program [16]. *Recall* is used as the metric to observe how many active-sites are identified outright by a program. Recall values are the number of sites identified by the program that are confirmed in SitesBase divided by the total number of binding sites more than two complete residues for given proteins in SitesBase; thresholds that are geared towards BsFinder in fact [16]. Since these results are readily reported, establish BsFinder as superior, and are conducted using techniques different than ours, we focused our approach on the centralized case study using the described target proteins. This approach takes advantage of the outlined target qualities described and address why msTALI can be discussed and used as our method for addressing successful alignments [1].

Though, BsFinder has some similarities [16], its output differs. First, while reporting the active-site results and confirmed active-sites it uses results consistent with the SitesBase database [24]. To remain consistent with our findings we will use PDB [22] directly to maintain our common core for confirmed active-site locations. Additionally, our reported active-site locations are measured in observance of our conserved regions, based on the number of amino acids in the protein (residues) [17]. With such we employ the following equations and our approach using msTALI for our precision and recall evaluating. From the definitions we have used msTALI in conjunction with annotations from PDB [22] to generate our results. Equation one, Eq 1, below, quantifies precision.

$$Precision = ASm_C / ASm_M \quad (1)$$

Here, $ASm$ refers to the actual number of active-sites obtained from the method; in our case msTALI. The subscript, $C$ denotes returned active-sites from the method that are confirmed as active-sites (confirmed by PDB for our approach). The subscript $M$ denotes actives-sites that are simply measured and outputted by the method.

Next, from the definitions we use Equation two, Eq 2, below, to quantify recall, which incorporates Eq. 3.

$$Recall = MSp - Error\ Score \quad (2)$$

$$Error\ Score = \frac{(ASm_M - ASm_C)}{ASg_C} * \varepsilon \quad (3)$$

Here, *MSp* refers to the maximum number of active-sites for a protein that can be achieved, which is essentially 100; for 100% of sites that can be described. We define our error score as the penalty evaluated from our precision. Whereby, the $\varepsilon$ multiplication addresses how great of a penalty factor we allot for. Therefore, $ASg_C$ is the number of active-sites confirmed from the ground truth comparison program, which again, we use annotations directly from PDB [22]. We then apply our scoring factor to obtain our actual recall values.

Literature reports the initial analysis of the target proteins by comparing the computational tools BsFinder, SiteEngine [14], SuMo [12], and pbdFun [15] for active-site analysis [16]. BsFinder is coined superior based on its recall and precision values. BsFinder shows an average recall of 82% while the highest amongst the other preexisting methodologies were less than 50%. With precision, BsFinder reports 34% while the other methods have precisions no higher than 21% [16]. Clearly, the objective is to have the highest percentage in these regards, and there is room for improvement.

## III. RESULTS AND DISCUSSION

Results of msTALI alignment are reported in this section. These results have been obtained by using the web version of the msTALI that can be found on: http://ifestos.cse.sc.edu/mstali. The use of this service is free and only requires an initial user registration.

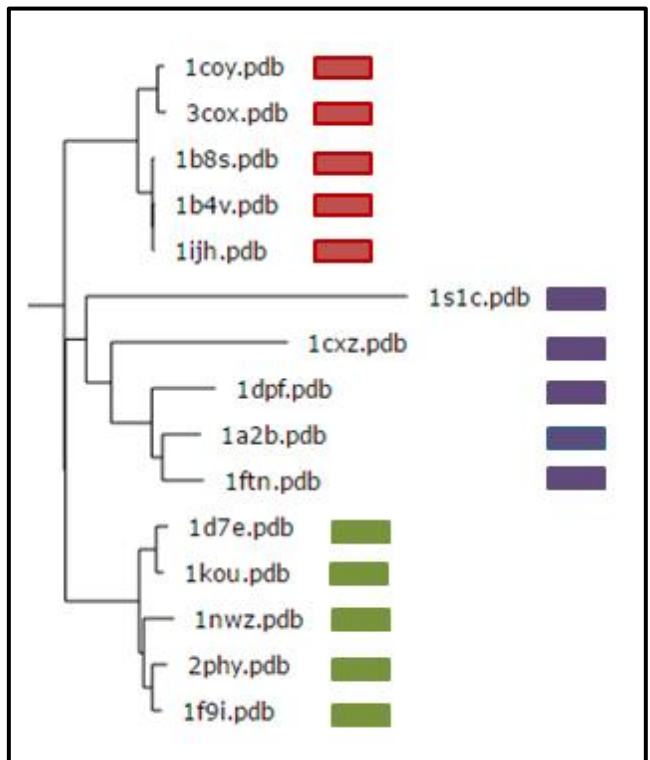

*Figure 2.* Phylogeny tree generated by using msTALI for a structural alignment on the 15 observed proteins simultaneously. Proteins marked with the same color illustrate the reconstruction of our subset grouping of fold families studied.

### A. Using msTALI phylogenetic analysis for subset validation

Prior to investigating the output generated by our multiple structure alignment with msTALI it is important to establish the legitimacy of its alignment [1]. The phylogenetic analysis feature of msTALI and its significance has been previously reported [17]. It is important to note that the phylogenetic analysis of msTALI is based on all encompassing features of the proteins including sequence, structure, and other biochemical properties [17]. To clarify further, by excluding all the information except the protein sequences, the results of msTALI will converge to that of ClustalW [25]. Fig. 2 conveys the annotated phylogenetic results of the msTALI analysis for the complete 15 protein target set. In this figure, the G proteins family in P-loop folds are colored purple, the PYP-like family in Profilin-like folds are colored green, and FAD-linked reductases family in FAD/NAD(P)-binding folds are colored red. As it can be observed in this figure, the clustering reconstructs the initial subsets accurately for each family. Additionally, the intermittent clustering of branches (though less important than the overall grouping) are consistent even when proteins may have multiple CATH [23] classifications (refer to Table 2). The success of msTALI in proper clustering of the proteins serves as a proof of concept to demonstrate the utility of msTALI in identifying the enzymatic activity of an unknown protein based on this method of clustering.

*Table 3.* Comparing the Precision and Recall for msTALI and BsFinder Amongst the Target Proteins. Results from BsFinder were previously reported [16] and used for comparison. *a. The first number is the number of output sites reported by the program (conserved regions in our case), the second number is confirmed sites from the program *b. and *c. The first number is the precision value (%), the second number is the recall value (%) for msTALI and BsFinder respectively.

|      | msTALI   |            | BsFinder  |            |
|------|----------|------------|-----------|------------|
|      | Number*[a] | Ratio (%)*[b] | Number*[a] | Ratio (%)*[c] |
| **1A2B** | 57\|16 | 28\|75 | 7601\|3647 | 48\|95 |
| **1CXZ** | 16\|11 | 69\|97 | 7832\|3602 | 46\|98 |
| **1DPF** | 37\|13 | 35\|83 | 7537\|3241 | 43\|92 |
| **1FTN** | 37\|14 | 38\|85 | 7435\|3121 | 42\|91 |
| **1S1C** | 37\|14 | 38\|87 | 7995\|3827 | 47\|99 |
| **1D7E** | 28\|6 | 21\|82 | 4845\|834 | 17\|58 |
| **1F9I** | 39\|9 | 23\|67 | 5771\|1068 | 18\|64 |
| **1KOU** | 37\|11 | 30\|84 | 5352\|1297 | 24\|59 |
| **1NWZ** | 28\|6 | 21\|80 | 5027\|1279 | 25\|63 |
| **2PHY** | 28\|6 | 21\|76 | 5451\|1189 | 21\|57 |
| **1B4V** | 37\|21 | 57\|95 | 7835\|4138 | 54\|97 |
| **1B8S** | 37\|21 | 57\|95 | 7996\|4101 | 52\|96 |
| **1COY** | 16\|7 | 44\|98 | 7892\|4135 | 53\|96 |
| **1IJH** | 165\|33 | 20\|59 | 7859\|4119 | 53\|96 |
| **3COX** | 37\|17 | 46\|93 | 7878\|4199 | 54\|98 |

*Table 4.* Comparison of four discussed programs as reported from the previous study observing 55 set of proteins [16].

| Program | Precision | Recall |
|---------|-----------|--------|
| msTALI | 37% | 84% |
| BsFinder | 34% | 82% |
| SiteEngine | 21% | 47% |
| SuMo | 11% | 25% |
| pdbFun | 15% | 11% |

### B. Precision and Recall analysis with msTALI

Our initial evaluation of msTALI focuses on simultaneous alignment of proteins based on the three subsets of fold families. This exercise is necessary to establish its pragmatic application in classification of unknown proteins. As described in section *2C*, previous investigations [16] report findings based on the precision and recall across the three subsets.

With our approach we get results from msTALI by aligning the five proteins from each group and based off of the relationships shown in our phylogeny tree in section *3A* In other words, msTALI results are obtained with our simultaneous alignments and any subset continuous runs [1]. All alignments are performed using the msTALI core/ default settings. We then utilize the conserved amino acid residues from our msTALI structural alignments to address precision and recall by confirming PDB [22] annotations. Reiterating, the ratios are obtained from the active-sites found by a program. They are then confirmed as active-sites for each protein by primary reference, and are used to calculate precision and recall, with aims to accurately recover 100% of proteins active-site locations.

Our overall precision and recall results are shown in Table 3. Here we explicitly list the results for each of the 15 target proteins using the conventions observed amongst all of our column one protein tables. Column two and three for Table 3 focus on the current results from this study as it pertains to msTALI. Columns four and five do the same using the preexisting results for the respective proteins, as referenced with BsFinder [16]. Column two of Table 3 list two numbers. The first number is the best resulting set of conserved amino acid regions resulting from an msTALI alignment. The second number is the corresponding number of returned conserved amino acid regions from msTALI that are confirmed as active-sites. Therefore, the values in column two are used directly with Eq. 1 for precision score. Further, column three first lists the precision value we mentioned, followed by the second number that is the calculated recall value for each protein using Eq. 2 for msTALI. Columns four and five, again, list the compared information for BsFinder. However, they show the methods and calculations previously conducted as it pertains to BsFinder comparison study and the details from section *2C,* which we report.

Noticeably, the values listed in columns two and four of Table 3 is substantially different. We will start by mentioning that msTALI reports based off of conserved numbers of amino acid residues that are then verified using PDB [22]. On the other hand, column four, with BsFinder report a representation of atoms verified by the SitesBase database [24]. Clearly, we will revisit this especially since there are a large number of atoms reported from the previous results [16].

To remain consistent with the previous work, we also report the performance of msTALI in identification of active-site in metrics of precision and recall related to the compared approaches. Here we aggregated the average precision and recall for msTALI across the 15 target set of proteins and included it to Table 4.

From Table 4 it is evident that msTALI has an average precision and recall higher than the existing methods. Our average precision is 37% for the 15 target protein set, and the msTALI recall is 84%. We are comfortable comparing our results on the 15 proteins to the previously observed methods which averaged on a study of 55 proteins. This is a direct result of having carefully used a target set of proteins that is difficult for msTALI with respect to active-site studies. Elaborating, we understand that we'll span an entire active-site space for proteins the more alignments we perform. This is attributed to our successful structural alignments [1]. However, some of our recall can be linked to overfitting. Overfitting, that becomes increasingly evident when proteins are strongly similar in structure as mentioned section *2B*. This is an issue common to many approaches that output too many matched atoms [16] and explains exactly why this is a challenging set of proteins for msTALI. Still our results report confirmed residues, and we address this issue in practice by performing alignments across multiple subsets during a target study; the three families, and then all 15 target proteins collectively name a few sets within this example. We account for overfitting by subtracting an outlier score from the maximum/ ideal recall of 100%. This validates our findings, and opens our discussion on comparing our approaches representation, precision, and why our results are reliable.

*Table 5.* Confirmed Protein length. The length of all 15 target proteins as annotated from PDB. We measure both the length of the protein and its active-site by amino acid residues and its length in atoms.

| PROTEIN | Length (Res) | Length (Atm) | Active Site Size (Res) | Active Site Size (Atm) |
|---|---|---|---|---|
| 1A2B | 182 | 1418 | 16 | 113 |
| 1CXZ | 182 | 2127 | 17 | 141 |
| 1DPF | 180 | 1400 | 14 | 94 |
| 1FTN | 193 | 1406 | 15 | 111 |
| 1S1C | 183 | 1411 | 18 | 126 |
| 1D7E | 122 | 943 | 12 | 106 |
| 1F9I | 125 | 989 | 9 | 79 |
| 1KOU | 125 | 944 | 16 | 133 |
| 1NWZ | 125 | 1135 | 11 | 107 |
| 2PHY | 125 | 1012 | 9 | 80 |
| 1B4V | 504 | 3849 | 33 | 236 |
| 1B8S | 504 | 3845 | 33 | 236 |
| 1COY | 507 | 3772 | 36 | 252 |
| 1IJH | 504 | 3901 | 32 | 228 |
| 3COX | 507 | 3739 | 29 | 203 |

Earlier, we mentioned how using PDB [22] to record protein size and confirmed active-sites is different from using confirmed representations from SitesBase [24]. We use Table 5 to translate our protein size to the number of atoms for both the protein and its active-site size. Columns three and five of the table report the confirmed corresponding length in atoms. Our initial reporting in amino acid length is still shown in columns two and four. Now we can use Table 3 and document the particular instances where msTALI stands out to BsFinder; since, at first glance they are comparable. It is documented that our largest proteins do not exceed 3900 atoms in length, and the largest documented active-site size does not exceed 260 atoms. Henceforth, any reported values that exceed these thresholds use a representation subject to overfitting.

We mentioned how overfitting is difficult for all methodology, but further for msTALI based on strong structural similarities. For this reason we have highlighted regions where msTALI does well considering the similarity. From Table 3, we highlight when msTALI outperforms BsFinder in yellow, with respect to precision. We highlight examples in green for recall. Values colored grey highlight msTALI results that are less than that of BsFinder but at a margin less than 5%. We observe that for the three fold families' the msTALI approach is more precise and surely has a higher recall identifying active-sites for protein from the PYP-like family in Profilin-like fold family. This makes sense because the fold family has a high amount of structural similarity, but the smallest length. Thus, the Profilin-like fold family is the most ideal scenario for our more challenging examples of using msTALI for active-site analysis.

Close analysis of Table 3, elicits additional points which illustrate the dissimilarity in measurements and approaches that exist between our msTALI and BsFinder. However, the aforementioned highlights the effectiveness of msTALI in the following ways: superior structural alignment, precision, and more important – since these findings are new with respect to the metric – recall. We use Table 3 and Table 5 in discussion on how our approach with msTALI is more reliable as compared to prominent used methods since it is consistent with PDB [22]. All annotated information is inline and when comparable, the difference between msTALI and BsFinder is three percent and eight percent for G proteins family in P-loop folds and the FAD-linked reductases family in FAD/NAD(P)-binding folds respectively, when BsFinder is seemingly useful. We consider this minor anomalies negligible, hitting home that our results are reliably reproducible and use a common standard from PDB [22] applicable to a multitude of applications.

### C. Active-Site identification with msTALI

It is evident that the intersection for conserved regions obtained from the three studies is difficult to analyze and that a simple intersection is not enough to deem them active-sites [1]. Our approach with msTALI successfully addresses these concerns. Though this study evaluates precision and recall, our overall purpose is active-site identification. In brief, we illustrate the conserved core region for the simultaneous alignment of all 15 proteins simultaneously. We provide examples from each of the fold families of this study, further verifying using msTALI as a suitable procedure for active-site description [1].

First, we use msTALI to test each of our three families. The G proteins family in P-loop folds had a conserved core of 173 residues, the PYP-like family in Profilin-like folds 119, and FAD-linked reductases family in FAD/NAD(P)-binding folds returned 495 residues. Considering the length of each target proteins with any of the three families, it is clear that results are consistent with the previous study, but also that msTALI does extremely well with our alignments. Nonetheless, additional observations need to be made to support our conclusion.

Next we evaluate subsets and partial categories from the outlined families. For example, we evaluate the alignment of a simple removal of one protein from each set and align them simultaneously. This results in an additional subset alignment of twelve proteins with a conserved core of 37 residues. Potential overfitting is addressed, and we continue analysis. Again, one could assume the following: if we have removed only one protein from each set for study, why not use a single protein from each family and use that conserved core region? In the case of a simultaneous alignment on one protein from G proteins family in P-loop folds, a protein from the PYP-like family in Profilin-like folds, and another form FAD-linked reductases family in FAD/NAD(P)-binding folds; 16 conserved core residues were returned from msTALI (results of a 3 protein subset inclusion). Considering our all-inclusive simultaneous alignment returned 28 residues, the 3 set alignments are valuable, but insufficient. Thus, to map the whole active-site space, and confirm our accuracy using PDB the most valuable results require the subsets of each alignment as depicted when addressing precision and recall.

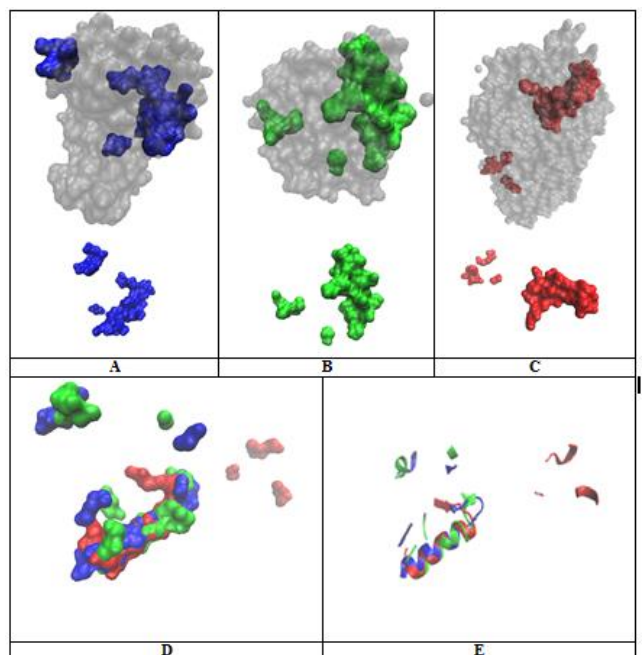

*Figure 3.* The conserved core regions observable from three proteins from each fold family, under the all-inclusive simultaneous alignment. (A) Illustrates conserved core residues for protein 1A2B as obtained from msTALI. (B) Illustrates conserved core residues for protein 1D7E as obtained from msTALI. (C) Illustrates conserved core residues for protein 1B4V as obtained from msTALI. (D) Illustrates the conserved core regions superimposed with respect to one another. (E) Depicts the Secondary Structural region described in D.

Fig. 3 depicts the conserved regions with respect to three of the proteins. Each serves as an abstraction to describe characteristics observable across the three families. *Section A.* illustrates the conserved region for protein 1A2B from the G proteins family in P-loop folds, *section B* does the same but for protein 1D7E from the PYP-like family in Profilin-like folds, and *section C* is protein 1B4V from the FAD-linked reductases family in FAD/NAD (P)-binding folds. *Section D* is a surface representation which superimposes the three conserved core regions from sections A, B, and C to convey the structural similarity for the motif common to both the complete target set and the three abstracted examples figured. Further, *section E* renders the same image picture in D, but with respect to secondary structure. The depicted conserved core regions from msTALI are consistent across all target proteins. Confirmed active-site regions are surface accessible or at the center of cavity/ cleft locations respective to protein family, and are located at the coil, non-structure, or turn and bend regions at the beginning of the alpha helical region of each target proteins conserved residues. Collectively, we observe that our simultaneous alignment on all 15 proteins yield precise results, characterizes motifs common to all the proteins observed, and endorses the validity in active-site bindings unique to each protein for studies outlined herein [26] and moving forward.

IV. CONCLUSION

Though the results of our methods are relatively comparable to the existing methods by exhibiting drawbacks with some fold families within the target set, our approach is ahead 73% of the time with respect to our target set (11 of 15 times). This explains why our overall average precision and recall exceeds the existing methods for this particular study from Table 3. We leverage that the target set is an outlier example for our comparison. This is evident since we observe (from table 3) that the 15 target proteins with the leading existing method fair higher than its overall reported precision and recall averages across its total study on 55 proteins. In our case we are specifically referring to protein examples that are difficult to our approach due to strong structural similarity. With such, we are in fact competitively outperforming existing methods. We obtained average precision using msTALI of 37% for our target set and 84% for recall on a set of proteins that we deem difficult for our approach.

Results further yield twenty-eight conserved residues across a simultaneous run on the targeted set of proteins. Additionally, all 15 proteins had locations within the conserved regions consistent with biologically confirmed residues mentioned by PDB for active-sites. Our phylogenetic analysis yielded a tree with annotations consistent with the three fold families represented, and these details are confirmed using CATH classification information. This solidifies our purpose of study and demonstrates the legitimacy of our application with biologically confirmed annotations. Future investigations will lead to exploration of even more classes of structures.

ACKNOWLEDGMENTS

This work was supported by NIH grant number P20 RR-016461 to Dr. Homayoun Valafar.